\documentclass[twocolumn,aps,prl,epsf,floats]{revtex4}

\begin{document}

{\bf Comment on "A Tale of Two Theories: Quantum Griffiths Effects in
Metallic Systems" by A. H. Castro-Neto and B. A. Jones:}

In a recent paper\cite{Neto04} Castro-Neto and Jones argue that
because  (as previously noted \cite{Neto00,Millis02}) the
observability of quantum Griffiths-McCoy effects in metals is
controlled by non-universal quantities, the quantum Griffiths-McCoy
scenario is a viable explanation for the non-fermi-liquid behavior
observed in heavy fermion compounds. Here we show, expanding on results of
 Ref.~\cite{Millis02},  that the "non-universal quantity" is 
simply the damping of the spin
dynamics by the metallic electrons. Quantum Griffiths-McCoy behavior
is only obtained when the damping is parametrically weak relative to
other scales in the problem, in other words when the spins are
essentially decoupled from the metallic electrons. This suggests
that in heavy fermion materials, where the Kondo effect leads to a
strong carrier-spin coupling, quantum Griffiths-McCoy effects are
unlikely to occur.

Quantum Griffiths-McCoy divergences arise from quantum mechanical
tunneling of disorder-induced spin droplets. For example, the
Griffiths-McCoy contribution to the magnetic susceptibility $\chi(T)$
may be expressed as
\begin{equation}
\chi(T)=\sum_N\frac{P(N)}{\omega_{tun}(N)+T}
\label{chi}
\end{equation}
Here $N$ is a measure of the droplet size and $P(N)\propto N^{-3/2}\exp \left( -N/N_{\xi }\right) $,
\begin{eqnarray}
\omega _{{\rm tun}}\left( N\right) &=&\left\{
\begin{array}{cc}
\omega _{0}\exp \left( -(\nu N_{c})\frac{N/N_{c}}{1-N/N_{c}}\right)  & N\leq
N_{c} \\
0 & N>N_{c}%
\end{array}%
\right.     \label{tunn}
\end{eqnarray}
are respectively the probability of finding a droplet of
size $N$ and the rate at which a typical droplet of size $N$ tunnels.

The essential feature of these equations is the  critical
droplet size, $N_c$, above which tunnelling ceases. In an insulator with
Ising symmetry one expects
$N_c=\infty$. Quantum Griffiths-McCoy behavior with exponent controlled
by $N(\xi)/\nu$ occurs as a balance
between the exponentially decreasing probability of finding a large droplet
and the exponentially increasing tunnelling time $\omega_{tun}^{-1}$.
However, in a metallic
system, the spin excitations of the particle-hole continuum typically
lead to an overdamped spin dynamics and thus to a finite, damping-dependent
$N_c$. Frozen droplets with $N>N_{c}$ give a
superparamagnetic $1/T$ contribution to $\chi $, whereas those with $N\leq
N_{c}$  may in appropriate
circumstances give a contribution to $\chi $ which
in some temperature regime is approximately a power
law divergence with exponent between $0$ and $1$ and which is larger
than the superparamagnetic contribution.

Eqs.~(\ref{chi}), (\ref{tunn}) and much of the physical picture were
given in Ref.~\cite{Neto00}; similar results were derived from a
different starting point in Ref.~\cite{Millis02}. Ref.~\cite{Neto04}
presents a detailed correspondence between the two approaches,
expanding on the comparison noted in Ref.~\cite{Millis02}.  The key
points, not discussed in \cite{Neto00,Neto04} are that the
'appropriate circumstances' in which quantum Griffiths-McCoy
behavior may occur are that $N_{c}$ is large enough that almost all
relevant droplets can tunnel, that the distribution of tunnelling
rates is appropriately broad and is  nearly continuous. We show that
this requires that the damping is very weak relative to the other
scales in the problem \cite{note2}.

A spin system embedded in a metal has, at sufficiently long times,
an overdamped dynamics characterized by a damping coefficient
$\Gamma$ which is expected to be of the order of the other scales in
the problem (such as the Fermi energy). We denote these scales
generically as $E_0$.  The conventions are such that $\Gamma \gg E_0$
corresponds to very weak damping and $\Gamma \ll E_0$ to very strong
damping (see Ref.~\cite{Millis02} for details).  Eqs.(3) and (4)
of Ref.~\cite{Neto04} show that $N_c \approx \Gamma/E_0$; thus
only for systems with parametrically small damping can $N_c$ be
large enough to allow a quantum Griffiths-McCoy regime.
Ref.~\cite{Neto00} obtained a large value for $N_c$ by using an
inconsistent procudure: the $\Gamma$ was calculated for the single
impurity Kondo problem (i.e. a 'light electron' value was used) but
an $E_0$ calculated for a dense Kondo system. A consistent
calculation  \cite{Millis04} for a dense Kondo system leads to
$\Gamma/E_0 \sim 1$

An additional constraint having to do with the tunnelling rate
distribution arises from a detailed analysis, which shows that for
quantum Griffiths-McCoy behavior to be observable a certain constant
$C_2$ must be much greater than unity (see Eqs.(50) of
Ref.~\cite{Millis02} or Eqs.(11) of Ref.~\cite{Neto04}).
Ref.~\cite{Neto04} uses an incorrect result for $C_2$. The correct
result is (see Eqs.(29)-(34) of Ref.~\cite{Millis02})
\begin{equation}
C_{2}=2\left( 1+\ln \left( c_{d}\frac{\Gamma \xi _{0}^{2}}{c^{2}\tau _{m}}%
\right) \right)
\end{equation}%
with $\ln c_{d}=0.115$ and $\xi_0$ a 'microscopic' length scale
expected to be of the order of a lattice constant.
Ref.~\cite{Millis02} found $\tau _{m}^{-1}=c^{2}/\left( \Gamma \xi
_{0}^{2}\right) $ implying that $C_{2}\simeq 2.2$ is a fixed number
of order unity and therefore not a number large enough to yield
significant quantum Griffiths-McCoy behavior. This estimate of
$\tau_{m}$ was derived on the assumption that the damping scale is
comparable to other scales in the problem. If $\tau_{m}$ is
determined by some damping-independent physics and if the damping is
parametrically weak ($\Gamma $ exponentially large) then a large
value of $C_{2}$ could occur. Thus we see again that, as expected,
one obtains quantum Griffiths-McCoy behavior only if the dissipative
contribution to the dynamics becomes unimportant.

To summarize, quantum Griffiths-McCoy behavior may occur in
disordered magnets with undamped or very weakly damped spin
dynamics, but not in magnets in which the damping is significant.
The essential feature of heavy fermion materials is a strong
carrier-spin coupling. While it seems very unlikely that in these
materials the damping is weak enough to allow quantum
Griffiths-McCoy divergences, a final resolution of this issue can
only come from experiments. To this end, we will show in a future
paper \cite{Millis04} how nuclear magnetic resonance and neutron
scattering measurements may be used to determine the ratio
$\Gamma/E_0$.

We thank A. Castro-Neto and B. Jones for stimulating
discussions. This work was supported by NSF-DMR-0431350 (AJM) and the Ames
Laboratory, operated for the U.S. Department of Energy by Iowa State
University under Contract No. W-7405-Eng-82 (J. S.).

\noindent A. J. Millis,

{\footnotesize Department of Physics, Columbia University}

{\footnotesize 538 West 120th Street, New York, NY 10027}

\noindent D. K. Morr

{\footnotesize Department of Physics, University of Illinois at Chicago,}

{\footnotesize Chicago, IL }

\noindent J. Schmalian

{\footnotesize Department of Physics and Astronomy and }

{\footnotesize Ames Laboratory, Iowa State University, Ames, IA 50011}


\begin{references}
\bibitem{note} The expression quoted in Ref.~\cite{Neto00} did appear in a
draft version of Ref.~\cite{Millis02}, but the error was corrected
in the published article.
\bibitem{note2} Even in this case, at sufficiently low temperatures the $1/T$
contribution from the frozen droplets will dominate the
susceptibility. For this reason we find Fig.~1 of Ref.~\cite{Neto04}
to be misleading: it discusses the low temperature limit but does
not show the 'frozen droplet contribution, focussing instead only on
terms which are subleading in this limit.
\end{references}
\end{document}